# A dressed polarizability framework for interface-coupled meta-atoms and large-scale metasurfaces

Peng Fu[1*], Jean-Paul Hugonin[2], Maxime Bertrand[3], Kevin Vynck[4], and P. Lalanne[1*]

[1] Laboratoire Photonique, Numérique et Nanosciences (LP2N), IOGS- Université de Bordeaux-CNRS, 33400 Talence cedex, France

[2] Laboratoire Charles Fabry (LCF, UMR 8501), CNRS−IOGS−Université Paris-Sud, 2 Avenue Augustin Fresnel, 91127 Palaiseau, France

[3] SharpEye SAS, 25 Avenue Pasteur, 94250 Gentilly, France

[4] Institut Lumière Matière (iLM), Université Claude Bernard Lyon 1, CNRS, Villeurbanne, France

* philippe.lalanne@institutoptique.fr

**Abstract.** The modeling of large collections of interacting meta-atoms remains a central challenge in photonics because it requires the simultaneous treatment of complex scatterer geometries, multiple scattering, and heterogeneous environments. Here, we introduce the dressed Global Polarizability Matrix (dGPM), a reduced-order electromagnetic framework in which complex meta-atoms are represented by compact scattering operators that explicitly incorporate the influence of nearby interfaces. The dGPM accurately reconstructs near- and far-field electromagnetic responses for interface-coupled structures, including geometries beyond the practical applicability of conventional T-matrix approaches. The framework naturally combines interface-coupled and free-space scatterers within a unified multiple-scattering formalism and enables simulations of large ordered and disordered metasurfaces. More broadly, the dGPM extends operator-based electromagnetic modeling to heterogeneous photonic environments while preserving a compact scatterer-based representation, providing an efficient framework for the simulation and design of large-scale photonic systems.

## 1. Introduction

Optical metasurfaces transform how we generate, shape, sense, and even perceive light [1]. They emerged in parallel with metamaterials in the 1990s [2,3] and, like their volumetric counterparts, rely on subwavelength resonant building blocks—commonly referred to as meta-atoms—whose individual and collective electromagnetic responses determine the macroscopic optical functionality of the device.

Two major classes of metasurfaces have emerged: ordered metasurfaces [4,5], in which meta-atoms are arranged on regular lattices, and disordered metasurfaces [6], where meta-atoms are distributed stochastically, generally subject to prescribed spatial correlations. In both cases, fabrication capabilities are advancing rapidly through complementary top-down lithographic and bottom-up self-assembly approaches [7,8]. As a result, the primary bottleneck is progressively shifting from fabrication itself toward the accurate simulation, inverse design, and optimization of increasingly complex metasurface architectures [9–12].

Despite sharing a common physical foundation as ensembles of spatially separated scatterers interacting through electromagnetic fields, ordered and disordered metasurfaces are typically modeled using markedly different numerical frameworks. Ordered metasurfaces are most often investigated with full-wave discretization techniques such as finite-difference time-domain (FDTD) methods [13] and rigorous coupled-wave analysis (RCWA) [14], whereas disordered metasurfaces are commonly analyzed using multiple-scattering formalisms, including the T-matrix method [15,16] and the discrete dipole approximation (DDA) [17–19].

Multiple-scattering approaches exploit the particulate nature of metasurfaces by discretizing only the scatterers while treating the surrounding medium analytically. Depending on the formulation, electromagnetic interactions are mediated through electromagnetic Green's functions, as in dipole-based methods, or through spherical vector wave expansions, as in T-matrix techniques. By avoiding volumetric discretization of free space, these methods often provide favorable computational scaling together with direct physical insight into collective multiple-scattering phenomena. Full-wave discretization techniques, in contrast, offer greater geometric and material generality at the cost of a substantially larger computational burden.

Yet the distinction between ordered and disordered metasurfaces is structural rather than physical. Both systems consist of interacting scatterers governed by the same electromagnetic principles, differing only in the spatial organization of their constituents. There is therefore no fundamental reason to maintain separate modeling paradigms. A unified multiple-scattering framework capable of treating both ordered and disordered metasurfaces would be not only physically natural but also computationally advantageous.

Existing multiple-scattering methods were primarily developed for disordered media. They face important limitations when applied to ordered metasurfaces. The T-matrix method [20], currently the reference approach for large-scale multiple-scattering calculations, represents electromagnetic fields using vector spherical wave functions defined outside the smallest circumscribing sphere of each meta-atom. For the high-aspect-ratio dielectric pillars commonly employed in metasurfaces [6,21], this sphere can be significantly larger than the physical scatterer, leading to inefficient field representations and degraded computational performance. The limitation persists in recent extensions of the method [22,23].

The DDA circumvents this geometric constraint by discretizing the scatterers into interacting dipoles. However, accurately representing high-index dielectric resonators typically requires a large number of dipoles per meta-atom [19], substantially increasing the size of the resulting multiple-scattering problem. As a consequence, much of the computational advantage gained by avoiding volumetric discretization of the surrounding medium is lost when large and dense metasurface ensembles are considered.

The Global Polarizability Matrix (GPM) method [24] was introduced to overcome the traditional trade-off between geometric flexibility and computational efficiency. In the GPM framework, each meta-atom is represented by a compact non-local polarizability matrix that is retrieved for every distinct geometry from a reference database of full-wave simulations of the scattered near field under a set of independent illuminations.

A key strength of the GPM formalism is that the resulting operator can be freely translated and rotated in three-dimensional space, enabling the efficient modeling of ensembles of meta-atoms with arbitrary positions and orientations, similarly to the T-matrix approach. Furthermore, the method has recently been implemented within the GPU-enabled automatic-differentiation framework PyTorch [25], opening promising perspectives for large-scale simulations and inverse-design strategies.

However, the original GPM formulation is restricted to meta-atoms surrounded by a homogeneous medium (Fig. 1a) and therefore cannot accurately describe what is arguably the most important configuration in nanophotonics: meta-atoms touching or straddling one or several interfaces. Hereafter, we overcome this important limitation by introducing the dressed Global Polarizability Matrix (dGPM), see Fig. 1b.

The remainder of this paper is organized as follows. In Section 2, we establish the theoretical framework of the dGPM and clarify its relationship with the original GPM formalism. In Section 3, we validate the method on high-aspect-ratio meta-atoms straddling a dielectric interface, a class of geometries that remains challenging for current T-matrix implementations.

In Section 4.1, we investigate strongly coupled meta-atom dimers and demonstrate that the dGPM can be naturally incorporated into the standard multiple-scattering formalism through Green tensors while maintaining high accuracy. Section 4.2 shows that conventional GPM and dressed dGPM operators can be seamlessly combined within a single simulation framework. This capability enables the treatment of heterogeneous systems containing meta-atoms embedded in homogeneous media, deposited on substrates, or intersecting dielectric interfaces. In Section 4.3, we apply the framework to ensembles comprising several hundred of meta-atoms arranged in either ordered or disordered configurations.

Finally, Section 5 discusses the scope, limitations, and perspectives of the GPM–dGPM framework. Particular attention is given to its applicability to large collections of interacting meta-atoms and to its relationship with existing multiple-scattering approaches for the modeling of metasurfaces and disordered photonic media.

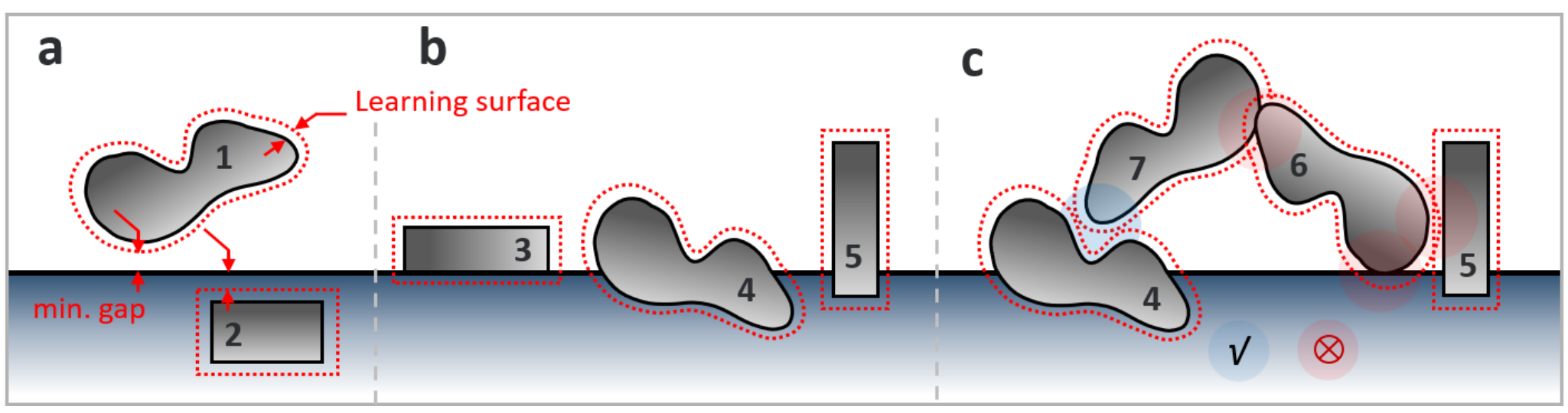

**FIG. 1. General framework for the GPM and dGPM for complex particulate systems. a,** The GPM is applicable only when the learning surface surrounding each meta-atom remains entirely within a homogeneous medium. Consequently, meta-atoms 1 and 2 can be described by the GPM. **b,** Meta-atoms deposited on, crossing, or embedded within an interface (meta-atoms 3–5) violate this condition and therefore require the dGPM formalism. **c,** Example of a heterogeneous configuration containing both standard and dressed operators. Meta-atoms 4–6 must be treated with the dGPM because their learning surfaces intersect the interface, whereas meta-atom 7 may be described using either the GPM or the dGPM. The multiple-scattering formalism remains valid only when the learning surfaces associated with different meta-atoms do not overlap. For example, the surfaces of meta-atoms 4 and 7 are non-intersecting, whereas those of meta-atoms 6 and 7 (or 5 and 6) overlap, thereby violating the assumptions of formalism.

## 2. Dressed Global Polarizability Matrix: Theory and Formulation

dGMP and GPM methods are conceptually very similar. So let us rapidly recap the main features of the GPM method, as they have been established in [24].

### 2.1. Global Polarizability Matrix

The GPM represents a meta-atom by a small set of $6N_p$ electric and magnetic numerical dipole components, which at first sight resembles the Discrete Dipole Approximation (DDA). The fundamental difference, however, lies in the interpretation of these dipoles and their associated polarizability matrix.

In the DDA, the dipoles represent physical polarization currents distributed throughout the scatterer. Each dipole is characterized by a local polarizability that relates its induced moment to the exciting field at the same location. As a consequence, the polarizability matrix is block-diagonal, and electromagnetic interactions between dipoles are introduced separately through the Green tensor appearing in the coupled-dipole equations.

The GPM follows a different philosophy. The numerical dipoles do not represent physical polarization currents and are not constrained to obey a local constitutive relation. Instead, the complete polarizability matrix is learned from reference scattering data. Both diagonal and off-diagonal blocks are therefore determined by numerical inversion and collectively encode the electromagnetic response of the entire meta-atom. In this representation, a given dipole component may depend on all $6N_p$ exciting field components acting on the numerical dipoles. The off-diagonal terms therefore capture internal electromagnetic couplings implicitly, without imposing any specific propagation law between the numerical dipoles.

This additional freedom allows complex meta-atoms to be represented with far fewer dipoles than required by the DDA while preserving an accurate description of their external electromagnetic response. In this respect, the GPM is conceptually closer to the T-matrix formalism than to conventional dipole discretization methods: a larger upfront computational effort is invested in constructing the operator, which can then be reused to efficiently solve multiple-scattering problems involving large collections of interacting meta-atoms.

The GPM operator is learned from reference data generated by independent full-wave simulations performed with a Maxwell solver (COMSOL is used throughout this work). Specifically, the near fields scattered by the meta-atom for a set of dipole sources positioned in its near field are computed on a learning surface, represented by the red dashed contour in Fig. 1a, which conformally surrounds the object. The GPM is then obtained through a least-squares inversion procedure from this large dataset. We refer to this stage as the *learning phase*.

To assess the predictive capability of the learned operator, we subsequently consider a different set of dipole sources from those used during learning and compare the fields predicted by the GPM with the corresponding full-wave simulations. We refer to this validation stage as the *generalization phase*. Successful generalization indicates that the scattering response of the meta-atom has been faithfully captured during learning, allowing the GPM to serve as a compact and accurate electromagnetic representation of the scatterer.

It is important to emphasize that the learned GPM is not an intrinsic material tensor associated with the meta-atom. Rather, it is a representation-dependent *effective scattering operator*. As such, it is not unique and depends on several modeling choices, including the positions of the numerical dipoles, the geometry of the learning surface, the excitation set used during learning, and the regularization strategy adopted during inversion.

A GPM is precomputed for a meta-atom embedded in a homogeneous medium and can thus be subsequently reused to model large ensembles of identical meta-atoms placed in stratified environments, provided that each meta-atom remains within the same background layer used during learning. Similar to a T-matrix, the GPM can be rigidly translated and rotated to account for translations and rotations of the corresponding meta-atom [24]. This portability is nevertheless conditional: the learning surface must remain entirely contained within a single homogeneous layer and must not intersect any interface, maintaining a finite gap from it, as illustrated in Fig. 1a.

One of the key advances introduced by the dGPM framework is the removal of this restriction. The dGPM remains applicable even when the learning surface intersects an interface, thereby enabling the treatment of meta-atoms deposited on, embedded within, or crossing material boundaries, as illustrated by meta-atoms 3–5 in Fig. 1b.

### 2.2 Dressed Global Polarizability Matrix

The dGPM method closely follows the GPM framework. The main difference is that the field scattered by the meta-atom is learned directly in the presence of the stratified medium. As a result, the dGPM is no longer an intrinsic property of the meta-atom but becomes dependent on the meta-atom environment. Any modification of the stratified medium therefore requires recomputing the dGPM. By contrast, the GPM can be reused for different stratified environments provided that the surrounding medium of the meta-atom remains unchanged.

The ability to handle learning surfaces that intersect material interfaces is gained at the expense of symmetry. Whereas the GPM supports arbitrary translations and rotations, the dGPM is restricted to translations parallel to the interface and rotations about the axis normal to the interface.

We now provide details of the implementation.

**Learning phase.** We start by generating reference data with COMSOL. A source surface (blue dashed line in Fig. 2) is defined at a distance $d_s$ from the meta-atom boundary, on which $N_s$ randomly distributed points are selected. At each point, six elementary electric and magnetic point sources are considered, and the field scattered by the meta-atom in its stratified environment is computed with COMSOL. This results in $6N_s$ independent excitation problems.

Next, a conformal learning surface (red dotted curve in Fig. 2) is introduced. On this surface, $N_l$ randomly distributed observation points are selected, where the six components of the scattered electromagnetic field are recorded. The scattered fields associated with all $(6N_s)$ excitations are assembled into a reference vector $\mathbf{\Psi}_s^{\mathrm{ref}}$ of size $36N_sN_l \times 1$.

We then compute the $6N_sN_d$ stratified Green-function matrices $G^{sd}$, which link the $6N_s$ source $\mathbf{S}$ to the $N_d$ numerical dipoles, using the formalism introduced in [24]. $G^{sd}$ is then used to evaluate the incident fields acting on the numerical dipoles, denoted symbolically by $G^{sd}\mathbf{S}$, and the corresponding induced electric–magnetic moments $\tilde{\mathcal{A}}G^{sd}\mathbf{S}$.

Similarly, we compute the $6N_lN_d$ stratified Green-function matrices $G^{dl}$, which link the numerical dipoles to the $N_l$ observation points on the learning surface. The scattered field predicted by the dGPM is then given by

$$\mathbf{\Psi}_s^{\mathrm{dGPM}} = G^{dl}\tilde{\mathcal{A}}G^{sd}\mathbf{S}, \tag{1}$$

from which the dressed polarizability matrix $\tilde{\mathcal{A}}$ is obtained by solving the regularized least-squares problem

$$\min_{\tilde{\mathcal{A}}} \ \| \mathbf{\Psi}_s^{\mathrm{dGPM}} - \mathbf{\Psi}_s^{\mathrm{ref}} \|^2. \tag{2}$$

The dGPM matrix retains the same block structure as the GPM,

$$\tilde{\mathbf{A}} = \begin{bmatrix} \tilde{\mathcal{A}}_{1\to1} & \tilde{\mathcal{A}}_{2\to1} & \cdots & \tilde{\mathcal{A}}_{N_d\to1} \\ \tilde{\mathcal{A}}_{1\to2} & \tilde{\mathcal{A}}_{2\to2} & \cdots & \tilde{\mathcal{A}}_{N_d\to2} \\ \vdots & \vdots & \ddots & \vdots \\ \tilde{\mathcal{A}}_{1\to N_d} & \tilde{\mathcal{A}}_{2\to N_d} & \cdots & \tilde{\mathcal{A}}_{N_d\to N_d} \end{bmatrix}, \tag{3}$$

where each block $\tilde{\mathcal{A}}_{m\to n}$ is a $6 \times 6$ tensor mapping the six incident field components driving numerical dipole $m$ to the 6 effective electric–magnetic moments induced at numerical dipole $n$. The off-diagonal blocks therefore capture the nonlocal response of the dressed numerical-dipole basis.

Unlike the GPM, for which the Green functions are evaluated in a homogeneous background medium, all Green functions appearing in Eq. (1) are computed in the stratified environment. They therefore include direct, reflected, transmitted, and evanescent contributions associated with the planar interfaces.

**Generalization phase.** To assess the predictive capabilities of the dGPM, a new set of randomly positioned sources is generated on the source surface together with a new set of randomly positioned observation points on the learning surface. The electromagnetic field scattered by the meta-atom is computed with COMSOL and assembled into a reference vector $\mathbf{\Psi}_s^{\mathrm{gen}}$. The corresponding dGPM prediction $\mathbf{\Psi}_s^{\mathrm{dGPM}}$ is obtained directly from Eq. (1).

The comparison is performed using the electromagnetic energy density $u_k = \frac{1}{2}[\varepsilon(\mathbf{r}_k)|\mathbf{E}(\mathbf{r}_k)|^2 + \mu(\mathbf{r}_k)|\mathbf{H}(\mathbf{r}_k)|^2]$ evaluated at each observation point $\mathbf{r}_k$ on the learning surfaces. This yields two vectors, $\mathbf{u}^{\mathrm{dGPM}}$ and $\mathbf{u}^{\mathrm{ref}}$, from which we compute the mean relative error and its standard deviation.

For two vectors $\mathbf{x}$ and $\mathbf{x}'$ of length $N$, the mean relative error as

$$\bar{E}(\mathbf{x},\mathbf{x'}) == \frac{1}{N}\sum_{k=1}^{N} E_k = \frac{1}{N}\sum_{k=1}^{N} \frac{|x_k - x'_k|}{|x_k| + |x'_k|}, \tag{4}$$

and the standard deviation $\sigma$ by $\sigma^2(\mathbf{x},\mathbf{x'}) = \frac{1}{N-1}\sum_{k=1}^{N}(E_k - \bar{E})^2$.

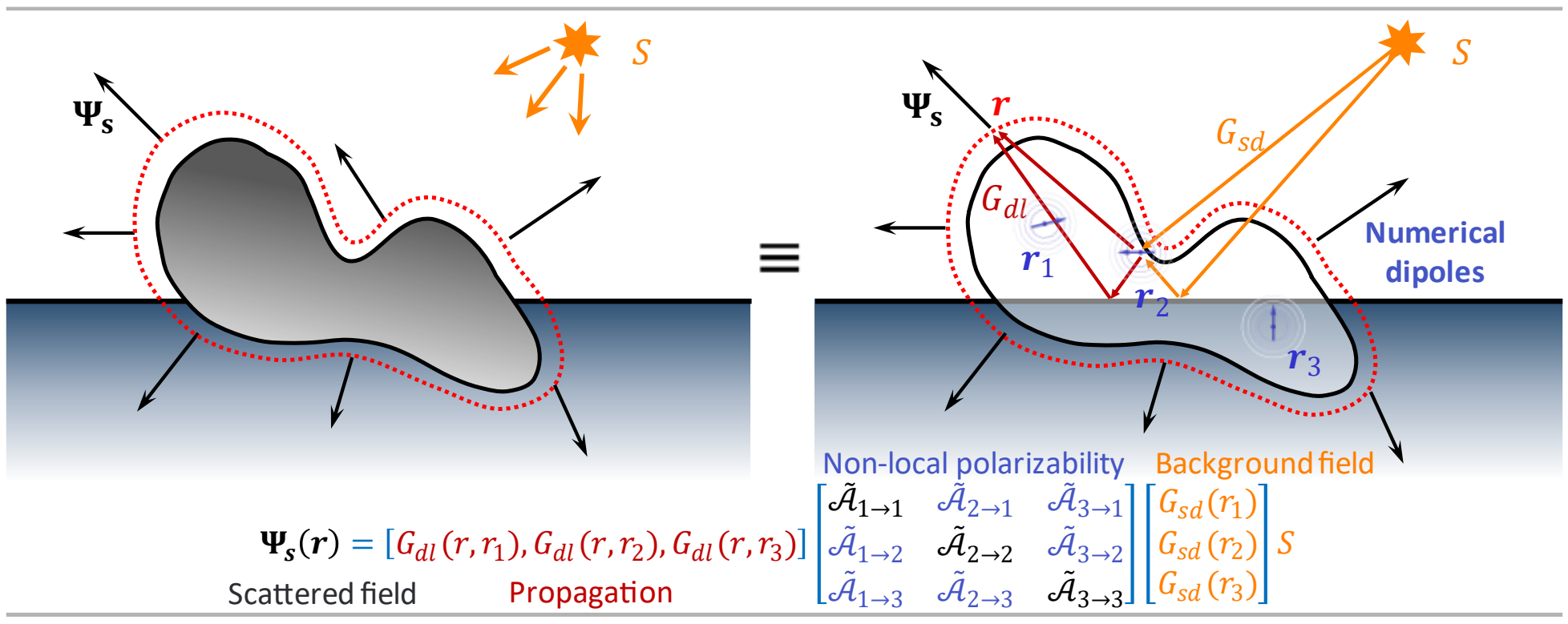

**FIG. 2. dGPM framework.** An arbitrary meta-atom in a stratified medium (left) is modeled by a small set of $N_p$ numerical dipoles (right) capable of accurately predicting the field scattered by the meta-atom. The figure shows the case $N_p = 3$. Each numerical polarizability is composed of 3 electric and 3 magnetic dipoles, and the dGPM denoted by $\tilde{\mathcal{A}}$ is composed of 3 × 3 subblocks. A concrete example of dGPM is given in Fig. 4b.

Although the dGPM framework provides a systematic route to constructing reduced-order models of meta-atoms in stratified environments, several practical aspects remain empirical. In particular, no universal threshold on $\bar{E}$ and $\sigma$ can be established to determine whether a given dGPM constitutes a sufficiently accurate electromagnetic representation of a meta-atom. Likewise, there is currently no general rule for selecting the number $N_d$ and locations of the numerical dipoles, nor for choosing the numbers of source and learning points, $N_s$ and $N_l$.

Nevertheless, our experience indicates that robust solutions for a broad range of meta-atoms [24] are obtained with typically the same parameters. The optimal numerical dipoles tend to be distributed along the principal symmetry axes of the meta-atom. For the pillar considered below, we use $N_d = 15$, $N_s = 30$ and $N_l = 1500$, leading to a mean relative error of $\bar{E} = 2.5\%$ and a standard deviation of $\sigma = 0.2\%$.

In practice, the learning procedure is repeated for $N_d \in \{1,2,\dots,N_d^{\max}\}$, yielding a family of operators $\tilde{\mathcal{A}}(N_d)$ together with their associated errors $\bar{E}(N_d)$. The computational cost required to generate this family is approximately 40 minutes for $N_d^{\max} = 25$.

### 2.3 Conceptual interpretation of the dressing

The term *dressed* is borrowed from condensed-matter physics, where the properties of a bare particle are renormalized by its interaction with the surrounding environment. In electromagnetism, a *dressed polarizability* similarly refers to the effective polarizability acquired by a scatterer through its interaction with neighboring scatterers or nearby interfaces (see Section 12.4.3 in [26]). This terminology naturally motivates the name *dressed Global Polarizability Matrix* (dGPM).

For meta-atoms whose learning surface does not intersect an interface, as in Figs. 1a and 3, both the GPM and dGPM frameworks can be applied. Although the two approaches may yield comparable accuracies, they differ fundamentally in their incorporation in the multiple scattering formalism.

To illustrate this distinction, consider the field driving a particular numerical dipole, for example the upper dipole of the rightmost meta-atom in Fig. 3. Several contributions enter this driving field. First, there is the incident field generated by the external source (orange arrows), which is identical in both formulations. Second, there is the field scattered by neighboring meta-atoms (blue arrows), which is also treated identically in both approaches.

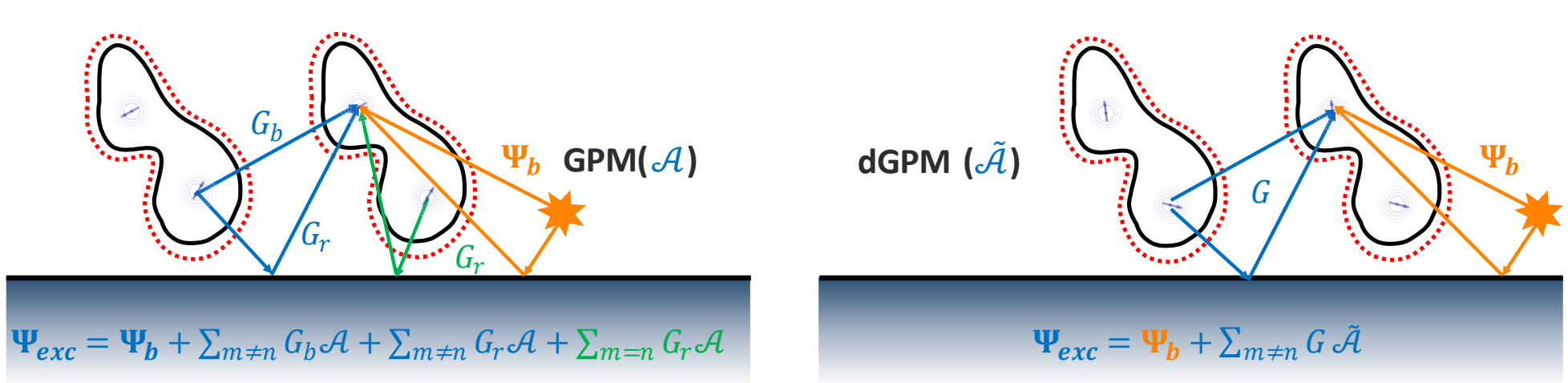

**FIG. 3. Multiple scattering in dGPM vs GPM frameworks.** In the bare GPM (left), the exciting field $\mathbf{\Psi}_{exc}$ involves three contributions: the incident field generated by the external source (orange arrows), the field scattered by neighboring meta-atoms (blue arrows), and the inter-scatterer substrate-mediated scattered field, including self-feedback (green arrows). In the dGPM formulation (right), these substrate-mediated interactions are incorporated into the dressed scattering operator. As a result, the exciting field contains only the incident field and the fields scattered by neighboring meta-atoms.

The key difference concerns the interaction between numerical dipoles belonging to the same meta-atom. In a layered environment, the Green function can be decomposed as $G = G_b + G_r$, where $G_b$ denotes the direct contribution that would exist in a homogeneous background medium and $G_r$ denotes the indirect contribution mediated by reflections and transmissions at the interfaces. Within the dGPM framework, both contributions are already embedded in the reference data $\mathbf{\Psi}_s^{\mathrm{ref}}$. Consequently, they are incorporated into the learned operator $\tilde{\mathcal{A}}$ through the optimization procedure of Eq. (2). The effect of the layered environment is therefore absorbed into the dressed polarizability matrix itself.

In contrast, GPM is learned in a homogeneous environment. Its off-diagonal blocks only capture the direct interaction described by $G_b$. The indirect contribution $G_r$, arising from the substrate-mediated interaction between numerical dipoles, is not contained in the learned operator and must instead be accounted for explicitly during the multiple-scattering calculation. This additional interaction is represented by the green arrow in the GPM schematic of Fig. 2c.

From this perspective, the dressing procedure can be interpreted as transferring part of the multiple-scattering physics from the Green function into the learned polarizability matrix. The dGPM thus incorporates substrate-mediated self-interactions directly into the meta-atom description, whereas the GPM treats them as external interactions that must be evaluated during the multiple-scattering calculation.

## 3. Validation of the dGPM for complex interface-straddling meta-atoms

In this section, we validate the dGPM framework using a test case deliberately chosen to lie beyond the range of applicability of conventional T-matrix approaches. Although the system consists of a single meta-atom, its geometry is nontrivial: it is composed of multiple materials and intersects a layered substrate, thereby preventing the definition of a closed learning surface entirely contained within a homogeneous medium.

The considered meta-atom is inspired by earlier work on metasurfaces for thermal infrared imaging at $\lambda = 10.6$ µm [27]. The composite structure is a ZnS-capped GaAs micropillar of total height $h = h_1 + h_2 = 6.8$ µm, standing on a ZnS-coated bulk substrate (Fig. 4a). It is obtained by etching a high-aspect-ratio GaAs pillar of diameter $D = 1$ µm and height $h_1 = 5.6$ µm into a bulk GaAs substrate ($\varepsilon_3 = 11.73$) and then depositing a low-index ZnS anti-reflection layer of thickness $h_2 = 1.2$ µm ($\varepsilon_2 = 4.84$) by electron-beam evaporation.

T-matrix methods—including extended formulations coupled with S-matrix approaches [23]—are not applicable to this configuration for two main reasons. First, the meta-atom straddles an interface, a situation that cannot be properly handled within the S-matrix framework. Second, when such high-aspect-ratio pillars are arranged in close proximity within a metasurface [27], the circumscribing spheres associated with neighboring meta-atoms overlap (blue dashed circles in Fig. 5a), causing breakdown of vector spherical harmonic expansions. In contrast, the dGPM circumvents these limitations entirely. Its learning surface conforms to the meta-atom geometry, allowing small separations between neighboring structures, and it imposes no homogeneity constraint, enabling the meta-atom to intersect multiple dielectric interfaces without restriction.

The dGPM is obtained through the numerical learning procedure introduced in Section 2, here adapted to the stratified geometry.

**Reference dataset.** A set of $N_s = 30$ dipolar sources is randomly distributed over a virtual source surface (dashed blue box in Fig. 2), located at a distance $d_s = 2$ µm from the meta-atom boundary. Each source is defined by six polarization states (three electric and three magnetic dipoles oriented along $x, y, z$), resulting in a total of $6N_s = 180$ full-wave training simulations, all performed using COMSOL Multiphysics with a very fine mesh to ensure negligible numerical discretization error in the reference data. For each simulation, the scattered field is sampled at $N_l \approx 1500$ points uniformly distributed over the conformal learning surface shown with dashed red box in Figs. 4a and 4c.

The total wall-clock time for the complete training pipeline is 64 minutes on a workstation equipped with an Intel Core i9 processor and 256 GB of RAM. or a rotationally symmetric meta-atom, the COMSOL simulations could be significantly accelerated by exploiting axisymmetry. This option was deliberately not used here; consequently, the reported runtime is representative of the computational cost associated with an arbitrary wavelength-scale meta-atom.

**dGPM computation.** The capped pillar is discretized into $N_d = 15$ numerical dipoles uniformly distributed along its longitudinal axis, yielding a $6N_d \times 6N_d$ dGPM $\tilde{\mathcal{A}}$, shown as a heatmap in Fig. 4b. The prominent off-diagonal blocks reveal strong spatial nonlocality, indicating significant coupling between all numerical dipoles. The learned dGPM achieves a relative energy-density error (Eq. 4) of $\bar{E}_{tra}$= 2.47% ($\sigma_{tra} = 0.27\%$) on the reference dataset.

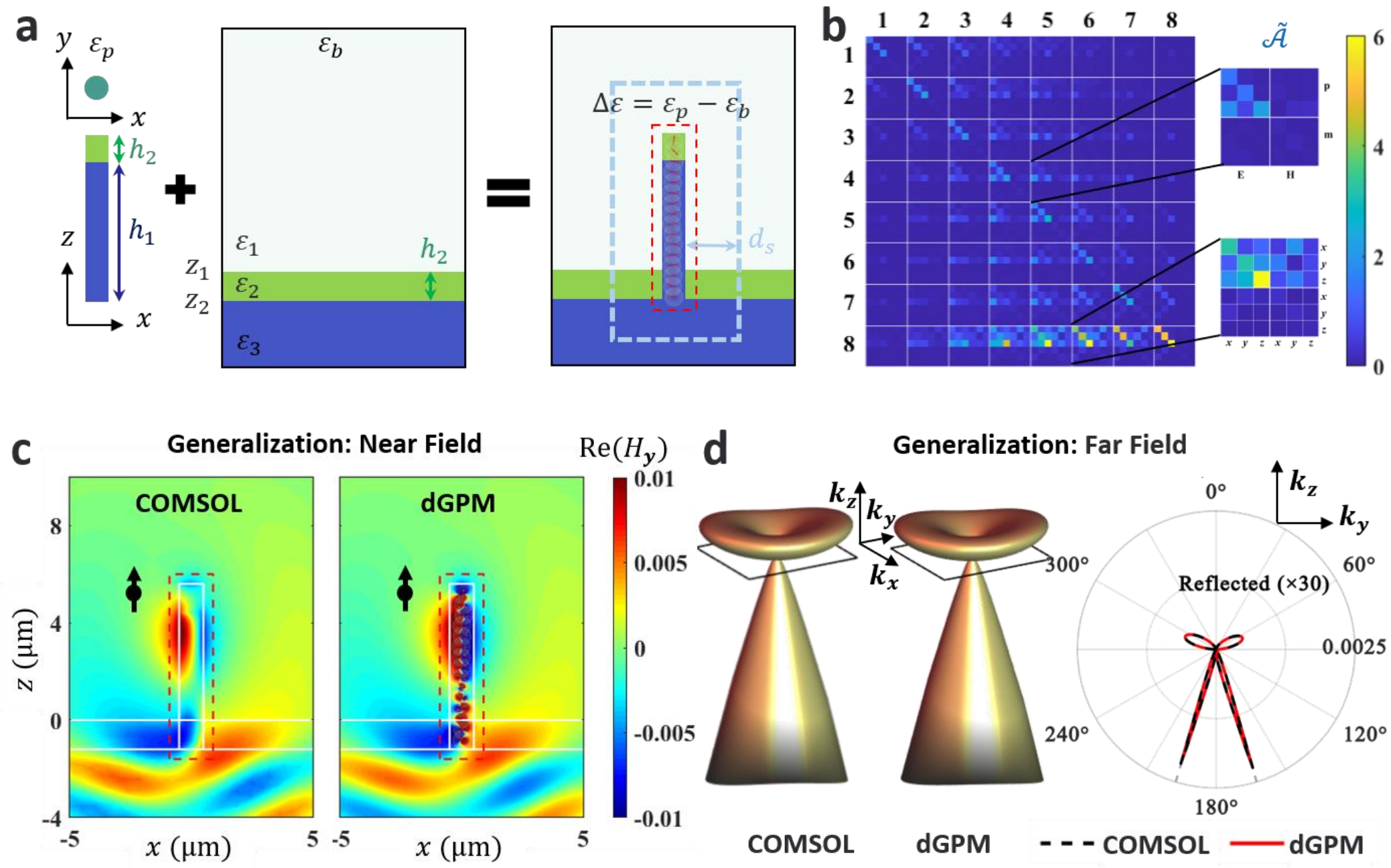

**FIG. 4. Validation of the dGPM for a complex interface-coupled meta-atom beyond the practical applicability of conventional T-matrix approaches. a**, ZnS-capped GaAs micropillar, with parameters $h_1 = 5.6$ µm, $h_2 = 1.2$ µm, $\varepsilon_3 = 11.73$ (GaAs), $\varepsilon_2 = 4.84$ (ZnS). The learning sources are distributed on a virtual surface (blue dashed box) at distance $d_s = 2$ µm from the high-aspect-ratio pillar. All tests are performed at $\lambda = 10.6$ µm. **b**, Heatmap of the dGPM $\tilde{\mathcal{A}}$ (illustrated here for $N_d = 8$ for simplicity). Each $6 \times 6$ block (see magnified insets)

describes the cross-coupling between electric ($p$) and magnetic ($m$) dipole moments. **c**, $Re(H_y)$ for an unlearned dipolar excitation shown with black arrows: COMSOL reference data (left) versus dGPM reconstruction (right). The dGPM accurately reproduces $Re(H_y)$ outside the learning surface (red dashed box), demonstrating generalization. **d,** Angular radiation pattern $dP/d\Omega$: comparison between COMSOL (black dashed) and dGPM (red). Excellent agreement is observed. Sharp features at the critical angle ($\theta_c = \text{asin}\,(\varepsilon_1/\varepsilon_3)^{1/2} \approx 16.96°$) capture the substrate light-cone effect.

**Generalization capability of the learned dGPM.** When evaluated on a generalization set of excitations not included in the training process, the relative error increases only marginally, from $\bar{E}_{tra}$ = 2.47% on the training set to $\bar{E}_{gen} = 2.72\%$ (standard deviation $\sigma_{gen} = 0.29\%$). This demonstrates that $\tilde{\mathcal{A}}$ does not merely constitute an overfitted encoding of the training data and effectively captures the scattering properties of the meta-atom, as well as its absorption since an accurate reconstruction of the scattered field also implies that the absorption is accurately predicted (Poynting theorem).

Figure 4c benchmarks the ability of the learned dGPM to reconstruct the near and far field for a specific source not included in the reference set, a $x$-polarized dipolar source at $[-2.36, 0.81, 5.22]$ µm (black arrows in Fig. 4c). It compares $|H_y|$ reconstructed with the dGPM with reference data computed with COMSOL. Inside the learning surface, as expected, the reconstruction is not accurate. In particular, field singularities appear near the numerical dipoles. This is an inherent feature of the point-dipole representation, which does not affect field reconstruction beyond the learning surface [24].

The mean relative error, Eq. (4), is only 3.96% for $H_y$ and 1.47% across all six electromagnetic components, ensuring that the vectorial nature of the scattering—including the phase and orientation of each field components—is accurately captured. This verification is essential, as it demonstrates that the reconstruction remains consistently accurate, even with the strong anisotropy induced by the dielectric interface.

The spatial diversity of the $N_d = 15$ numerical dipoles, distributed across three distinct material environments (ZnS in air, GaAs in air, and GaAs in ZnS), appears to play a key role in this performance. With dipoles within each of these environments, the dGPM captures the local polarization response of the different parts of the structure, while the non-local interaction matrix A self-consistently couples them through the stratified Green tensor. This combination of local sampling and non-local coupling enables the model to account for both the heterogeneous composition of the meta-atom and the abrupt change in background environment at the interface. As a result, the dGPM accurately resolves the complex scattering contributions arising from the different regions of the structure.

Figure 4d compares the far field radiation diagram obtained from the near fields computed with COMSOL or the dGPM, using the near-to-far-field-transform freeware RETOP [28]. The radiation pattern shows an intense forward-scattering profile consistent with high-efficiency anti-reflection [27] and directional antenna-like [29] behaviors. The transmitted energy forms a hollow-cone distribution, with sharp intensity maxima at the boundaries of the light cone imposed by the high-index GaAs substrate. Indeed, the near-field accuracy translates into the far field. To quantify the agreement between the two angular radiation maps, we compute the integrated absolute difference between the dGPM and COMSOL far-field intensities over all emission directions, normalized by the total radiated power obtained from COMSOL. This yields an error of 0.68%, confirming that the near-field accuracy of the dGPM is preserved after the near-to-far-field transformation.

The radiation pattern arises from two concurrent effects. First, the ($z$)-oriented incident electric dipole source, denoted ($P_z$), imposes radiation nulls along its dipole axis. Second, the substrate restricts propagating transmission to the critical-angle cone, $\theta_c = \text{asin}(\varepsilon_1/\varepsilon_3)^{1/2} \approx 16.96°$, represented with the grey dashed lines in Fig. 4d. Outside this cone, radiation arises from evanescent-to-propagating wave conversion at the substrate interface [29]. The dGPM accurately reproduces both the sharp angular cutoff and the characteristic null structure of the radiation pattern, demonstrating that $\tilde{\mathcal{A}}$ captures the complete evanescent-to-propagating coupling physics of the multilayer geometry.

## 4. Application to multi-meta-atom systems

We now consider a system of two interacting particles to validate the implementation of multiple scattering within the dGPM and GPM frameworks. Section 4.1 benchmarks the accuracy of inter-particle coupling in a two-pillar system and quantifies the benefits of the approach relative to T-matrix methods. Section 4.2 demonstrates the hybridization of GPM and dGPM operators within a single simulation. Section 4.3 presents a global validation of the approach for metasurfaces composed of several hundreds of meta-atoms, discusses the current limitations of the framework and outlines directions for future developments.

### 4.1 Pillar dimer

Using the dGPM trained in Section 3, we assemble two identical ZnS-capped GaAs pillars in a collinear configuration (Fig. 5a) and vary the side-to-side separation $d$ from 0.4 µm to 6.0 µm. The system is illuminated by a TE-polarized plane wave at wavelength $\lambda = 10.6$ µm with an in-plane incidence angle $\theta$.

The geometric advantage of the GPM framework over T-matrix methods is immediately apparent. The circumscribing sphere associated with each pillar has a radius $r_0 \approx 3.4$ µm. Consequently, for a two-pillar system, the spheres overlap for $d < 5.87$ µm, causing the T-matrix expansion to lose convergence throughout nearly the entire practically relevant separation range. In contrast, the conformal learning surfaces used by the dGPM overlap only for $d < 0.8$ µm, extending the accessible operating range by almost an order of magnitude. This advantage becomes even more pronounced for high-aspect-ratio structures, since as the pillar height increases, the T-matrix exclusion zone expands, whereas the dGPM validity limit remains primarily determined by the lateral footprint of the learning surface.

Figure 5b presents the reflection and transmission cross sections, ($\sigma_R$) and ($\sigma_T$), as functions of the incidence angle for separation distances ranging from 0.4 µm to 6.0 µm. The dGPM predictions (solid lines) are compared against COMSOL reference simulations (blue dots). To quantify the agreement, Table 1 reports the mean relative errors of the scattered powers averaged over all sampled incidence angles.

Three regimes can be identified. At large separations ($d = 6$ µm), the scattered power is approximately twice the scattered power of an isolated pillar, despite a residual electromagnetic interaction expected at $d \approx \lambda/2$ [30]. For intermediate separations, $d > 0.8$ µm, corresponding to the nominal validity range of the dGPM, the relative error averaged over all angular configurations remains below 1.8%. At $d = 0.6$ µm, the learning surfaces slightly overlap, but the relative errors remain moderate (0.9% in reflection and 2.2% in transmission), indicating a degree of robustness beyond the nominal validity limit. Only at the smallest separation considered, $d = 0.4$ µm, do the errors increase significantly (7.4% in reflection and 3.6% in transmission), marking the practical operating limit of the present framework for the chosen learning surface.

| $d$ (µm) | $d/\lambda$ | Within nominal validity domain | Reflection error (%) | Transmission error (%) | Status |
|---|---|---|---|---|---|
| 6.0 | 0.56 | yes | 1.1 | 0.67 | accurate |
| 2.0 | 0.2 | yes | 1.6 | 0.52 | accurate |
| 0.8 | 0.08 | yes | 0.4 | 1.8 | accurate |
| 0.6 | 0.06 | no | 0.9 | 2.2 | still accurate |
| 0.4 | 0.04 | no | 7.4 | 3.6 | inaccurate |

**TABLE 1.** Mean relative errors of the reflection and transmission intensities radiated for the two-pillar system at selected separation distances $d$. The dGPM remains accurate well within its nominal validity domain ($d > 0.8$ µm) and retains acceptable accuracy slightly beyond it ($d = 0.6$ µm). Significant deviations arise only at extreme proximity ($d = 0.4$ µm).

We now examine in detail the configuration $\theta = 30°$ and $d = 2$ µm (orange stars in Fig. 5b), which lies well within the nominal validity domain while exhibiting significant inter-pillar coupling. Figures 5c and 5d compare the near and far fields reconstructed by the dGPM with the corresponding COMSOL reference solutions. Outside the learning surfaces, the dominant electric-field component $E_x$ is virtually indistinguishable from the exact numerical solution. Notably, the dGPM accurately resolves the field enhancement within the narrow inter-pillar gap, a spatial regime that is fundamentally inaccessible to T-matrix methods.

The reconstructed electric-field amplitude $E_x$ exhibits a mean relative error (Eq. (4)) of 1.82% for the dominant component and 2.43% when averaged over all six electromagnetic field components. This near-field accuracy translates directly to the far field (Fig. 5d), where the reconstructed three-dimensional radiation pattern exhibits a global energy-distribution relative error of only (2.3%) relative to COMSOL. The dominant forward-scattering lobes and the strongly suppressed back-reflection are both accurately predicted with the dGPM. The two-dimensional polar cut further confirms the quantitative agreement in the angular interference pattern, demonstrating that the dGPM preserves the inter-particle phase relationships required for coherent multiple scattering.

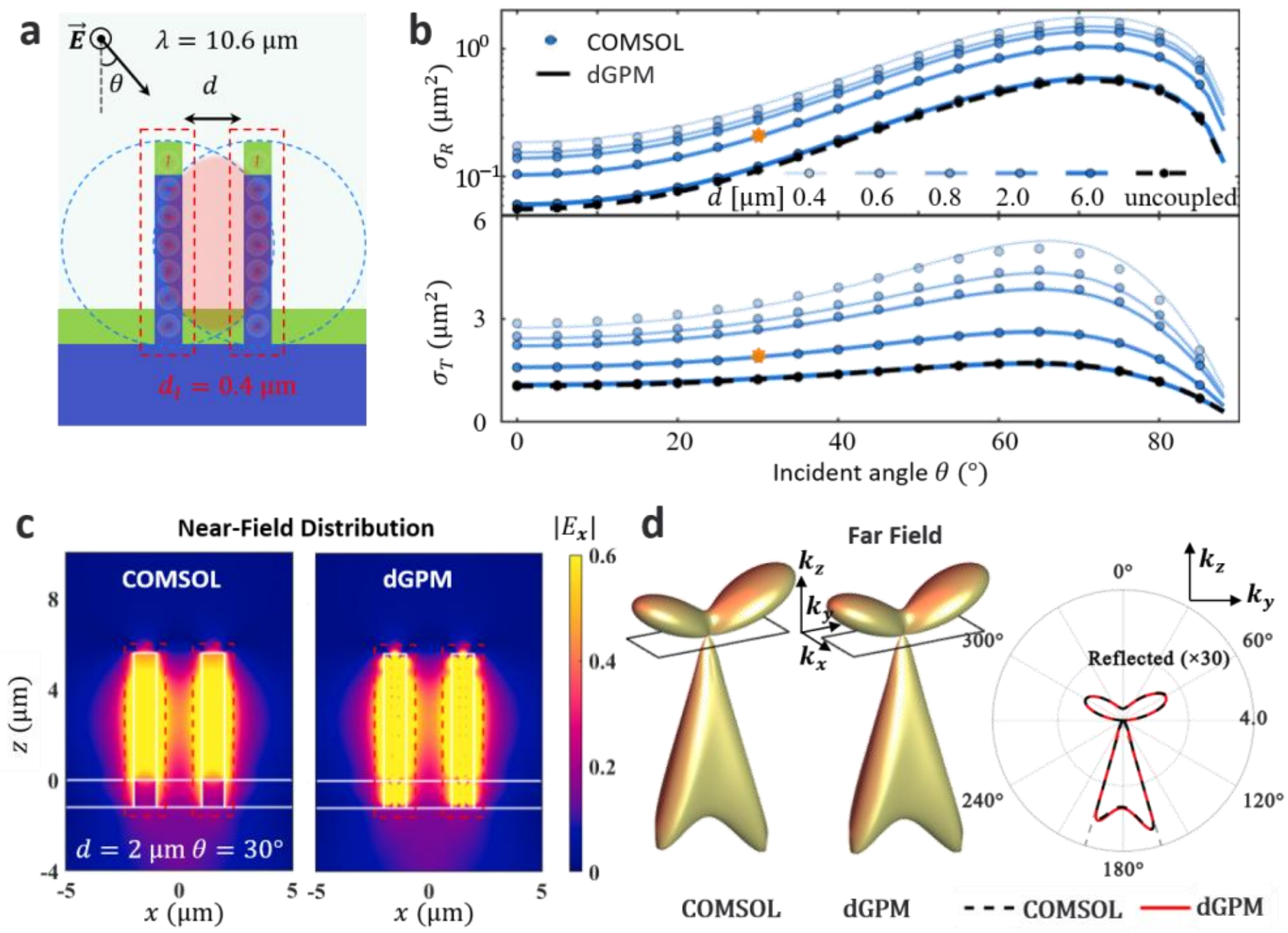

**FIG. 5. Validation of the dGPM for strongly coupled interface-coupled meta-atoms.** at $\lambda = 10.6$ µm. **a**, Two identical pillars separated by a side-to-side distance $d$ and illuminated by a TE-polarized plane wave incident at an angle $\theta$. Blue dashed circles denote the circumscribing spheres required in a conventional T-matrix, which becomes invalid when the spheres overlap for ($d < 5.9$ µm). In contrast, the dGPM remains applicable until the learning surfaces (red dashed lines) overlap, corresponding to much smaller separations ($d < 0.8$ µm). **b**, Reflection and transmission cross sections, $\sigma_R$ and $\sigma_T$, obtained by normalizing the radiated powers by the incident power flux, as functions of the incidence angle $\theta$ for several separation distances $d$. The dGPM accurately captures the evolution of the optical response from the strongly coupled regime to the uncoupled limit. For $d = 6$ µm, the latter (bold dashed line), corresponding to twice the scattering power of an isolated pillar, is recovered. Orange stars indicate the configuration ($\theta = 30°, d = 2.0$ µm) examined in **c** and **d**. **c**, Near-field reconstruction of $|E_x|$ for $\theta = 30°, d = 2.0$ µm. COMSOL reference solution (left) and dGPM reconstruction (right). Outside the learning surfaces, the dGPM accurately reproduces the reference field and captures the strong near-field interaction in the gap region, despite the high aspect ratio of the pillars and their close proximity. **d**, Corresponding far-field scattering patterns. For clarity, the amplitude of the reflected lobe is multiplied by a factor of 30.

For the same high precision, each COMSOL reference point for the pillar dimer in Fig. 5b requires nearly 12 minutes and 125 MB of RAM, whereas the dGPM evaluates each incident angle in approximately 1 second following a one-time initialization of 9 second and with a memory footprint of only 0.5 MB. This represents a 700-fold acceleration in speed and a 250-fold reduction in memory usage, enabling near-real-time parameter sweeps and statistical analysis.

### 4.2 Combined GPM-dGPM frameworks for complex systems

The GPM models meta-atoms embedded in a homogeneous medium and can be freely translated and rotated in 3D space, enabling the efficient modeling of ensembles with unrestricted positions and orientations [24]. In contrast, the dGPM has reduced symmetry due to the presence of interfaces: meta-atoms can only be translated within planes parallel to the substrate and rotated about axes perpendicular to the interface. Combining both approaches is therefore advantageous for systems containing both freely suspended meta-atoms in homogeneous regions and meta-atoms intersecting layered interfaces.

We demonstrate this capability on a three-body system consisting of a lossy cubic meta-atom (side length $L = 1\ \mu$m, refractive index $n = 2.1 - 0.6i$) buried beneath a pillar dimer ($d = 1$ μm) that straddles the air–substrate interface (Fig. 6a). We define $g$ as the vertical separation between the learning surface surrounding the cube and those associated with the surface pillars. For a cube learning-surface offset of $d_{l,cube} = 0.1$ μm and a pillar learning-surface offset of $d_{l,pillar} = 0.4$ μm extending below the interface, $g = 0$ corresponds to the two sampling manifolds being in direct contact.

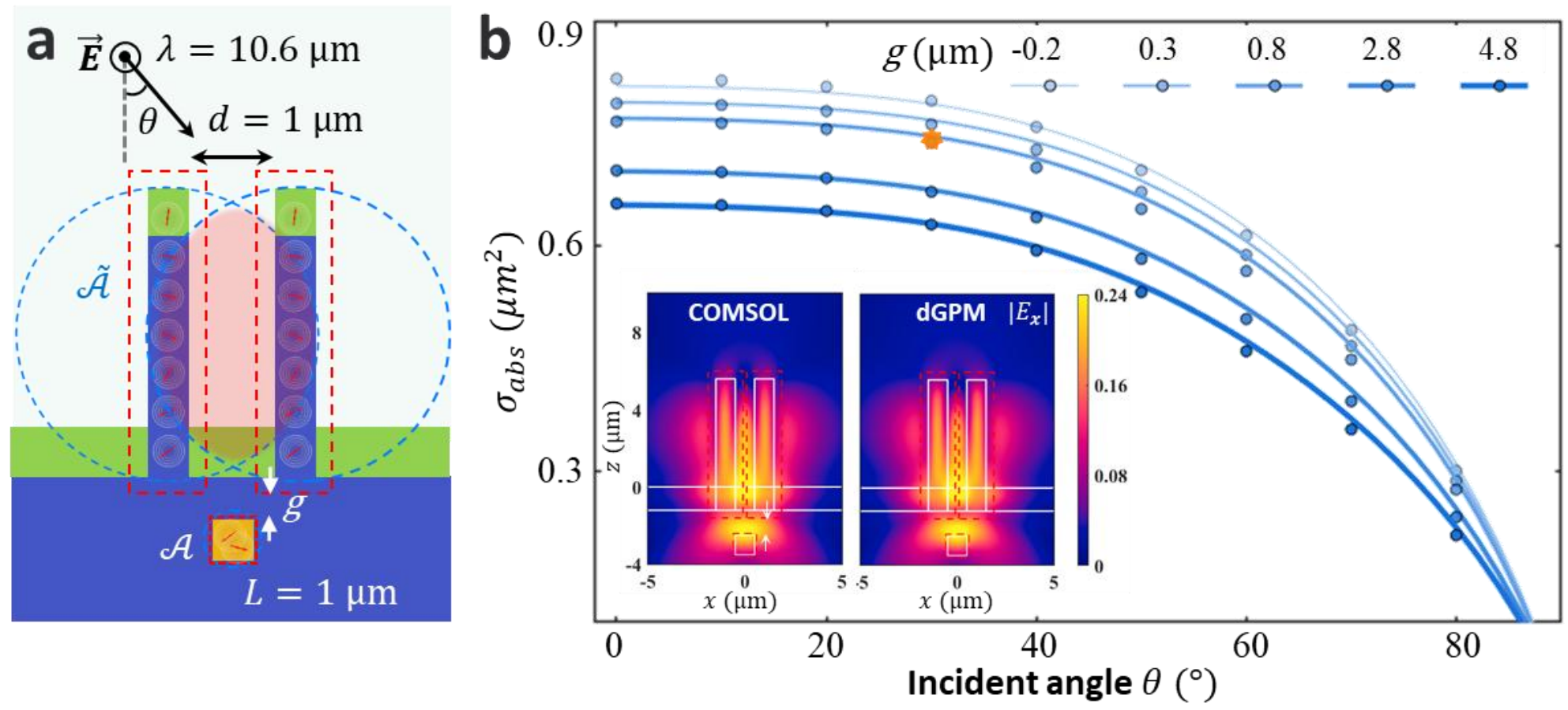

**FIG. 6. Unified operator-based modeling of heterogeneous photonic environments. a**, Schematic of a lossy GPM-cube (side length $L = 1\ \mu$m, refractive index $n = 2.1 - 0.6i$) buried beneath a dGPM-pillar dimer. The conformal learning surfaces are offset by 0.1 μm from the cube and 0.4 μm from the pillars. The parameter $g$ denotes the vertical separation between the respective learning surfaces: $g = 0$ implies contact. **b**, Absorption cross section, obtained by normalizing the absorbed power by the incident power flux, as a function of the incidence angle $\theta$ for various separation distances $g$. The hybrid GPM–dGPM framework (solid lines) exhibits quantitative agreement with COMSOL reference calculations (dots) for $g \geq 0.3$ μm and remains remarkably robust even when the learning surfaces overlap $g = -0.2$ μm. Inset: Near-field comparison of $|E_x|$ obtained with COMSOL and the combined framework for $\theta = 30°$ and $g = 0.8$ μm (orange star). Note that the buried cube could also be described using a dGPM operator. However, because the interface-mediated interactions vary with $g$, a separate dGPM would need to be constructed for each separation distance. This example therefore highlights the advantage of combining reusable GPM operators with environment-specific dGPM operators within a single simulation.

Figure 6b displays the absorption cross section $\sigma_{abs}$ of the buried cube as a function of the incidence angle $\theta$ for various separation distances $g$. The combined GPM–dGPM framework (solid lines) exhibits excellent agreement with COMSOL reference calculations (dots) for $g \geq$ 0.3 µm, faithfully predicting the complex interplay between depth-dependent reflections and multi-body coupling. Notably, the framework remains remarkably robust even for $g = -0.2$ µm, where the sampling manifolds overlap. Despite this violation of the nominal separation condition, the model continues to capture the correct absorption trends, with only minor quantitative deviations. This behavior highlights the intrinsic robustness of our operator-based formulation to moderate perturbations of the local near-field environment.

Importantly, the absorption associated with the buried meta-atom is not evaluated from an internal volumetric mesh. Instead, it is obtained from an energy-balance procedure based on the learned operator: the extinction is computed from the response of the GPM object, while the scattering contribution is evaluated from the radiated far field, so that the absorption cross section follows from: $\sigma_{abs} = \sigma_{ext} - \sigma_{sca}$. This provides an efficient way to evaluate dissipated power in complex environments, e.g. bidisperse mixture [31], although the numerical dipoles should not be interpreted as physical dipoles carrying a local material absorption.

### 4.3 Large ensembles

In this section, we assess the scalability of the dGPM for large meta-atom ensembles. Two representative configurations are considered: ordered arrays, in which many inter-particle separations are repeated, and disordered arrays, in which most separations are unique.

The computation consists of two stages. The first is a one-time initialization, during which the dense Green-interaction matrix is assembled for all numerical dipoles in the ensemble. The second is an illumination-dependent self-consistent solve. For a large number $N$ of meta-atoms, the initialization time scales approximately as $O(N^2)$, consistent with the pairwise construction of the interaction matrix. The solution time exhibits a steeper scaling, close to $O(N^3)$, as expected for the solution of a dense linear system.

Ordered and disordered ensembles display similar overall scaling behavior, indicating that the framework is not restricted to highly symmetric arrangements. The main difference arises during initialization. In ordered arrays, repeated pair separations allow identical Green-tensor evaluations to be reused, reducing the assembly cost. Consequently, the dense linear solve becomes the dominant computational expense for systems containing as few as ($N \approx 150$) meta-atoms. In disordered arrays, by contrast, most pair separations are distinct, limiting such reuse and shifting the crossover to larger systems, around ($N \approx 10^3$).

This behavior is consistent with the implementation, for which the most computationally intensive part of the initialization is the evaluation and assembly of the layered-medium Green tensor. As shown in the inset of Fig. 7, the memory footprint follows the expected dense-matrix scaling, approximately $O(N^2)$, for both ordered and disordered systems.

The reported timings should be regarded as indicative rather than optimized benchmarks. The current MATLAB implementation has not been optimized for high-performance computing and does not include explicit problem-specific parallelization; any multithreading arises from MATLAB and its underlying linear-algebra libraries. Profiler analysis shows that the runtime is mainly distributed between two operations: construction of the dense Green-interaction matrix and the dense self-consistent linear solve. For moderately sized disordered systems, the former remains dominant because most Green-tensor evaluations are unique. For sufficiently large arrays, however, the dense linear solve becomes the main bottleneck due to its faster scaling, especially in ordered configurations where Green-tensor reuse substantially reduces the initialization cost.

All timings were measured in MATLAB R2023b on a Dell Precision 7920 Tower workstation running Windows 10 Pro for Workstations, equipped with dual Intel Xeon Silver 4210R CPUs, 20 physical cores, 40 logical processors, and 256 GB RAM. MATLAB was allowed to use up to 20 computational threads. No GPU acceleration was used.

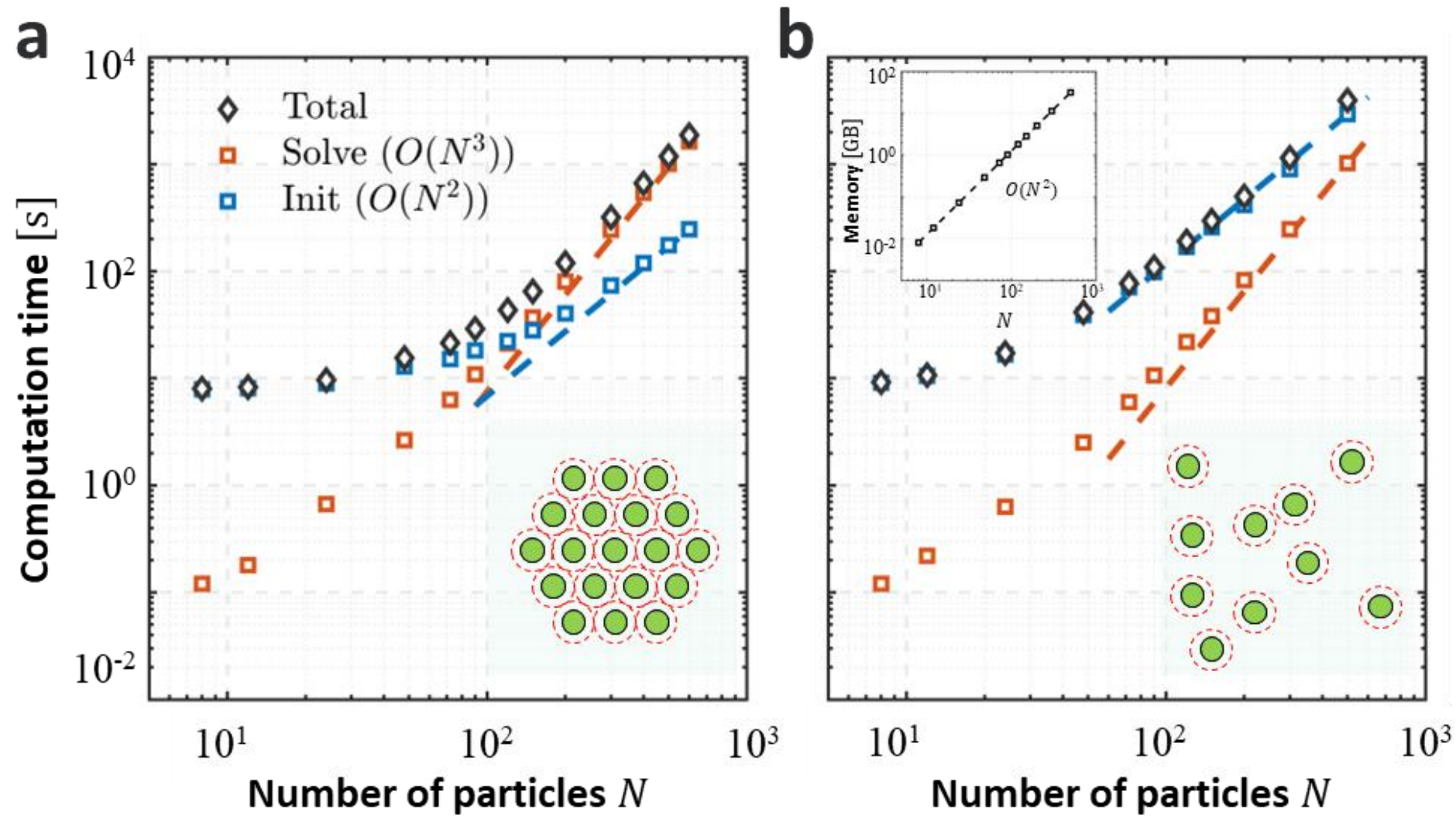

FIG. 7. **Computational scaling of the GPM–dGPM framework for large meta-atom ensembles.** Wall-clock initialization and solve times as functions of the number of interacting meta-atoms $N$ for **a,** ordered and **b,** disordered ensembles containing between 10 and 1000 scatterers. The initialization stage corresponds to the construction of the dense Green-interaction matrix, whereas the solve stage corresponds to the computation of the self-consistent multiple-scattering solution for a given illumination. Dashed lines indicate the expected $N^2$ and $N^3$ scalings associated with matrix construction and dense linear-algebra operations, respectively. The inset in **b** shows the memory footprint, which follows the expected $N^2$ scaling of a dense interaction matrix. These results demonstrate the applicability of the framework to large collections of interacting meta-atoms while identifying the dominant computational costs. Reported timings are indicative only, as the present MATLAB implementation is not optimized and does not exploit hardware acceleration or problem-specific parallelization.

Indeed, we cannot benchmark the results of Fig. 7 against finite-element simulations performed with our COMSOL software, as the corresponding computational requirements would exceed the memory available on our desktop workstations. Nevertheless, we expect the dGPM predictions to remain highly accurate even for large-scale metasurfaces. The only approximation introduced by the method lies in the dipolar representation of the meta-atoms, whose accuracy has been extensively validated and shown to introduce only small errors.

This behavior contrasts with conventional finite-element or finite-difference approaches, where the propagation medium must be discretized. In such methods, maintaining accuracy over increasingly large structures generally requires progressively finer meshes because phase-velocity errors accumulate over long propagation distances. By contrast, the dGPM represents wave propagation through Green functions, thereby preserving the exact continuous-space phase evolution between scatterers and avoiding discretization-induced numerical dispersion. As a result, the accuracy of the propagation model does not degrade as the system size increases.

## 5. Conclusion

We have introduced the dressed Global Polarizability Matrix (dGPM), a reduced-order electromagnetic framework for the modeling of complex meta-atoms interacting with dielectric interfaces. By incorporating Fresnel reflections and evanescent near-field feedback directly into the scattering operator, the dGPM extends operator-based electromagnetic modeling from homogeneous media to heterogeneous photonic environments. The resulting formalism enables the accurate reconstruction of near- and far-field responses for interface-coupled structures while preserving the compact representation that underpins the efficiency of the original GPM approach.

The framework was validated on representative infrared meta-atoms, including high-aspect-ratio pillars straddling dielectric interfaces and strongly coupled pillar dimers. In all investigated cases, the dGPM reproduced the reference electromagnetic response with relative errors of only a few percent. The method accurately captured both local field distributions and far-field observables, including scattering, extinction, and absorption, demonstrating that compact operator representations remain effective even in the presence of strong substrate-mediated interactions and near-field coupling. Importantly, the dGPM remained robust in challenging configurations involving moderate overlap of the learning surfaces, suggesting that its practical domain of validity extends beyond the formal assumptions used during its construction.

The method was validated on representative infrared metasurface building blocks, including high-aspect-ratio pillars intersecting a substrate and strongly coupled pillar dimers. In all investigated cases, the dGPM reproduced the reference electromagnetic response with relative errors below a few percent while maintaining a substantially reduced computational cost compared with full-wave simulations. For the pillar-dimer benchmark, the dGPM provided more than two orders of magnitude reduction in memory usage together with speedups approaching three orders of magnitude once the meta-atom operators had been constructed, compared to COMSOL.

A key feature of the framework is its compatibility with the original GPM formalism. By combining environment-independent GPM operators with interface-specific dGPM operators within a unified multiple-scattering formulation, heterogeneous photonic systems containing buried, suspended, and interface-coupled meta-atoms can be modeled self-consistently. This capability opens the door to the efficient simulation of complex ordered and disordered metasurfaces, as well as more general heterogeneous photonic environments.

Several challenges nevertheless remain. The optimal placement of numerical dipoles for arbitrarily shaped meta-atoms remains largely unexplored, and the construction of accurate operators still requires the generation of reference datasets. As recently demonstrated for T-matrix methods, shared libraries of precomputed operators for common geometries and environments could substantially reduce this effort and facilitate broader adoption of the approach [20].

The present implementation already enables simulations involving up to thousands of interacting meta-atoms. While the current MATLAB code exhibits the expected $O(N^2)$ memory scaling and $O(N^3)$ computational scaling associated with dense interaction matrices, these limitations are not intrinsic to the formalism. Significant improvements remain possible through optimized implementations, iterative solvers, GPU acceleration, and hierarchical matrix techniques.

More broadly, the dGPM demonstrates that compact operator-based descriptions can be extended to interface-coupled and heterogeneous photonic systems without sacrificing accuracy. We anticipate that this framework will provide a valuable complement to existing full-wave and multiple-scattering methods for the analysis, optimization, and large-scale simulation of metasurfaces and disordered photonic media.

Acknowledgements. The authors acknowledge Miao Chen, Minggang Luo, Denis Rideau from STMicroelectronics Grenoble and several other colleagues for helpful feedback during the preparation of the manuscript. PL acknowledges financial support from the European Research Council Advanced grant (Project UNSEEN No. 101097856) and from Bpifrance, IPCEI Electronique (Project COS No. DOS0218539/00)

Disclosures. The authors declare no conflicts of interest.